\documentclass[pra,10pt,twocolumn,superscriptaddress,showpacs]{revtex4}

\usepackage{epsfig}
\usepackage{amsmath}
\usepackage{graphicx}
\usepackage{graphics}
\usepackage{float}
\usepackage{amssymb}
\usepackage[usenames,dvipsnames]{color}
\usepackage{natbib}
\usepackage{epstopdf}
\usepackage{verbatim}
\usepackage{wrapfig}

\newcommand{\bra}[1]{\left\langle #1\right|}
\newcommand{\ket}[1]{\left| #1\right\rangle}

\newcommand{\jn}[1]{~{\bf #1}}
\newcommand{\opav}[3]{\left\langle #1 | #2 | #3 \right\rangle}

\newcommand{\beq}{\begin{equation}}
\newcommand{\eeq}{\end{equation}}

\begin{document}

\title{Role of initial system-environment correlations: A master equation approach}

\author{Adam Zaman Chaudhry}
\affiliation{NUS Graduate School for Integrative Sciences
and Engineering, Singapore 117597, Singapore}
\author{Jiangbin Gong}
\email{phygj@nus.edu.sg}
\affiliation{NUS Graduate School for Integrative Sciences
and Engineering, Singapore 117597, Singapore}
\affiliation{Department of Physics and Center for Computational
Science and Engineering, National University of Singapore, Singapore 117542,
Singapore}

\begin{abstract}
In order to achieve practical implementations of emerging quantum technologies, it is important to have a firm understanding of the dynamics of realistic quantum open systems. Master equations provide a widely used tool in this regard. In this work, we first construct a master equation, valid for weak system-environment coupling, which explicitly takes into account the impact of preparing an initial system state from an equilibrium system-environment state that has system-environment correlations. We then investigate the role of initial system-environment correlations using this master equation for a system consisting of many two-level atoms interacting with a common environment. We show that, in general, due to the initial system-environment correlations before a state preparation, the quantum state of the system can evolve at a faster time-scale.  Moreover, we also consider different initial state preparations, and demonstrate that the influence of state preparations depends on the initial states prepared. Our results can be of interest to many topics based on quantum open systems where system-environment correlation effects have been neglected before.

\end{abstract}

\date{\today}
\maketitle

\section{Introduction}

Any realistic quantum system is not closed - it is always interacting with its environment. Consequently, the system dynamics cannot be described by the Schrodinger equation, and in general, finding the reduced
system dynamics is a highly non-trivial problem. Various approaches have been developed, with perhaps the most common one being that of master equations \cite{breuerbook}. The basic idea is to consider the total system consisting of the system of interest and the environment as closed, which hence can be evolved using the usual unitary time evolution. The environment degrees of freedom can then be eliminated to obtain a differential equation that describes the system dynamics only.

It should be noted, however, that generally speaking, in order to obtain a master equation that is amenable to analytical or numerical solutions, various approximations and assumptions have to be invoked. For example, it is often assumed that the system-environment coupling is weak, and that the environment loses knowledge about the system state very quickly (the Markovian approximation). It is also commonly assumed that the initial system-environment state is a simple product state consisting of the initial state of the system and a thermal bath state for the environment. This assumption is usually justified on the grounds that, at least for weak coupling, the initial system-environment states should not play a significant role \cite{PollakPRE2008}. Moreover, for Markovian environments, the state of the environment cannot act as a `memory' for the system \cite{Modi2011}. Any effect of the initial correlations is then quickly lost.

It is known, however, that in many situations of current experimental research, these approximations cannot be made. For instance, for strong system-environment coupling, not only can a weak-coupling approximation not be made, but also the initial system-environment coupling can have a noticeable effect on the system dynamics \cite{PechukasPRL, RoyerPRL, Hanggi, TanimuraPRL2010, cheekong}. Due to this fact, as well as the increased interest in non-Markovian dynamics \cite{BreuerJPhysB2012, FrancononMarkovian}, the initial uncorrelated state assumption has come under close scrutiny recently, with various studies being performed to investigate its validity \cite{HakimPRA1985, HaakePRA1985, Grabert1988, SmithPRA1990, GrabertPRE1997, PazPRA1997, LutzPRA2003, BanerjeePRE2003, vanKampen2004, vega, ErezNature2008, GordonNJP2009, BanPRA2009, GordonNJP2010, UchiyamaPRA, SmirnePRA2010, DajkaPRA, TanPRA2011, MorozovPRA2012, gaoarxiv, SeminPRA2012, wgwang, ZhangPRA2010, ChaudhryPRA2013}. Most of the studies performed to date have considered single-body systems - a single spin or a single harmonic oscillator coupled to a thermal bath. There are, however, notable exceptions \cite{TanimuraPRL2010, ZhangPRA2010, ChaudhryPRA2013}. In particular, it has been found that if the systems consists of many two-levels systems (TLSs) coupled to a common environment, then the effect of the initial system-environment correlations can be enhanced depending on the number of particles in the system, even though each TLS may be weakly coupled to the environment \cite{ChaudhryPRA2013}. However, this study, and indeed almost all other studies of the effect of initial system-environment correlations, has been performed using an exactly solvable model, which are the exception rather than the rule \cite{breuerbook}.

It is the purpose of this work to go beyond exactly solvable models while taking into account initial system-environment correlation effects. To this end, we intend to construct a master equation, valid in the weak coupling regime, using specific initial state preparation with the initial system-environment correlations accounted for.  We then apply this master equation to a system of many TLSs coupled to a bath of harmonic oscillators, and we show that as we increase the number of TLSs in the system, the effect of initial correlations on the dynamics becomes significant.  This is because the environment state will be more affected by the initial state preparation as the number of TLSs increases.   Moreover, we investigate such an effect of initial correlations for different initial state preparations.

This paper is organized as follows. In Sec.~II, we introduce our master equation based on explicit initial state preparation, with the initial system-environment correlation before the state preparation accounted for. In Sec.~III, we apply this master equation to study the dynamics of a system of two-level atoms coupled to a common environment. We approximate the system dynamics at short times in Sec.~IV to show the significant influence of the initial correlations for a large number of two-level atoms. We consider different initial states in Sec.~V, and we summarize our results in Sec.~VI. Details of some calculations are given in the appendices.

\section{Formalism}

We first present a master equation to calculate the reduced system dynamics, starting from an initial state prepared with a projective measurement and with the initial system-environment correlations incorporated, which is correct to second-order in the system-environment coupling strength. The derivation presented here relies essentially on basic perturbation theory, and assumes little familiarity with the theory of open quantum systems. Alternatively, the same master equation can be derived using the time convolutionless approach, as we explain in Appendix \ref{TCLappendix}.

We begin by writing the total system-environment Hamiltonian as
\begin{equation}
H = H_S + H_B + \alpha V \equiv H_0 + \alpha V,
\end{equation}
where $\alpha$ is a parameter that keeps track of the order of the coupling strength between the system and the environment. At the end of the calculation, we will set $\alpha = 1$.
From first-order perturbation theory, we can write the system-environment unitary time-evolution operator as
\begin{equation}
\label{unitaryoperatorperturbative}
U(t) \approx U_0(t) \left[ 1 - i\alpha \int_0^t ds \widetilde{V}(s) \right],
\end{equation}
with $\widetilde{V}(s) = U_0^\dagger (s)V U_0(s)$, and $U_0(t)\equiv U_{S}(t)\times U_{B}(t)$ is the `free' unitary time evolution operator, that is, the time evolution operator corresponding to $H_0$.
We now note that
\begin{equation}
\rho_{mn}(t) = \text{Tr}_S \left[ Y_{nm} \rho(t) \right] \equiv \langle Y_{nm} \rangle,
\end{equation}
where $Y_{nm} = \ket{n}\bra{m}$, $\ket{n}$ and $\ket{m}$ being any basis states of the system, and $\rho(t)$ is the system density matrix at time $t$. This expression can be rearranged to give
\begin{align}
\langle Y_{nm} \rangle &= \text{Tr}_{\text{S,B}} \left[ (Y_{nm} \otimes 1_B) \rho_{\text{tot}}(t) \right], \notag \\
&= \text{Tr}_{\text{S,B}} \left[ U^\dagger (t) (Y_{nm} \otimes 1_B) U(t) \rho_{\text{tot}}(0) \right], \notag \\
&= \text{Tr}_{\text{S,B}} \left[ X_{nm}^H(t) \rho_{\text{tot}}(0) \right],
\label{eqY}
\end{align}
where the superscript $H$ denotes time evolution with $U(t)$, that is, $X^{H}_{nm}(t) \equiv U^\dagger(t) X_{nm} U(t)$. It follows that
\begin{equation}
\label{MEsource}
\frac{d\rho_{mn}(t)}{dt} = \text{Tr}_{\text{S,B}} \left[ \rho_{\text{tot}}(0) \frac{dX_{nm}^H(t)}{dt} \right].
\end{equation}
Our objective now is to derive a perturbative expression for $\frac{dX_{nm}^H(t)}{dt}$. First note that $X_{nm}^H(t)$ is a Heisenberg picture operator. As such, it obeys the Heisenberg equation of motion
\begin{equation}
\label{heisenbergequationofmotion}
\frac{dX_{nm}^H(t)}{dt} = i[H_0^H(t), X_{nm}^H(t)] + i[V^H(t), X_{nm}^H(t)].
\end{equation}
Now, using Eq.~\eqref{unitaryoperatorperturbative}, we can write
\begin{align}
X_{nm}^H(t) &= U^\dagger(t)X_{nm}U(t) \notag \\
&\approx \widetilde{X}_{nm}(t) + i\alpha\int_0^t ds [\widetilde{V}(s), \widetilde{X}_{nm}(t)],
\end{align}
where the tildes denote time evolution under $U_0(t)$. This means that $\widetilde{V}(t) = U_0^\dagger (t) V U_0(t)$.
Similarly,
\begin{equation}
V^H(t) \approx \widetilde{V}(t) + i\alpha \int_0^t ds [\widetilde{V}(s), \widetilde{V}(t)].
\end{equation}
By substituting these two expressions in Eq.~\eqref{heisenbergequationofmotion}, it can be shown that
\begin{align}
\label{allterms}
\frac{dX_{nm}^H(t)}{dt} &= i[H_0^H(t),X_{nm}^H(t)] + i\alpha[\widetilde{V}(t),\widetilde{X}_{nm}(t)] \notag \\
&+ \alpha^2 \int_0^t ds  [[\widetilde{V}(t),\widetilde{X}_{nm}(t)],\widetilde{V}(s)],
\end{align}
Given an initial condition, by substituting Eq.~\eqref{allterms} in Eq.~\eqref{MEsource} we can derive a master equation. Usually, this task is performed using the initial state
\begin{equation}
\label{nocorrinitialstate}
\rho^d_{\text{tot}}(0) = \rho(0) \otimes \rho_B,
\end{equation}
with $\rho_B = e^{-\beta H_B}/Z_B$ and $Z_B = \text{Tr}_B [e^{-\beta H_B}]$. How can this state come about? We allow the system and the environment to come to equilibrium. Then, if the system and the environment are interacting with vanishing interaction strength, the total thermal state is
\begin{equation}
\rho_{\text{tot}} = \frac{e^{-\beta H}}{Z} = \frac{e^{-\beta H_S}}{Z_S} \frac{e^{-\beta H_B}}{Z_B},
\end{equation}
where $Z_S = \text{Tr}_S [e^{-\beta H_S}]$. This is because $\alpha \rightarrow 0$. Due to the vanishing coupling strength, the states of the system and the environment are uncorrelated. We can then perform a selective projective measurement at $t = 0$, described by the projector $\ket{\psi}\bra{\psi}$, to prepare the system in the state $\ket{\psi}$. Since the system and environment are uncorrelated, this measurement affects only the system.
We then obtain the initial system-environment state given by Eq.~\eqref{nocorrinitialstate}.

We can now ask what happens for finite coupling strength. In this case, the system-environment state before the projective measurement is given by
\begin{equation}
\rho_{\text{tot}} = \frac{e^{-\beta H}}{Z_{\text{tot}}}.
\end{equation}
This is the equilibrium system-environment state, and due to the finite system-environment coupling strength, it cannot, in general, be factorized into a system part and an environment part - this state is correlated \cite{ErezNature2008, GordonNJP2009, GordonNJP2010, cheekong}. The corresponding system equilibrium state is $\rho = \text{Tr}_B[\rho_{\text{tot}}]$ and the environment equilibrium state is $\rho_B = \text{Tr}_S[\rho_{\text{tot}}]$. It should be noted that $\rho_B$ is now, in general, not the thermal bath state $e^{-\beta H_B}/Z_B$.

We next perform a selective projective measurement on the system alone. This time, because the system and the environment are correlated, we obtain
\begin{equation}
\label{withcorrinitialstate}
\rho_{\text{tot}}(0) = \ket{\psi}\bra{\psi} \otimes \dfrac{\opav{\psi}{e^{-\beta H}}{\psi}}{Z},
\end{equation}
where $Z$ is the normalization factor such that $\text{Tr}_{\text{S,B}}[\rho_{\text{tot}}] = 1$. The measurement has the effect of removing the correlations between the system and the environment. The total system-environment state is then no longer in equilibrium. Note that the environment state is different from the thermal bath state $e^{-\beta H_B}/Z_B$ for two reasons: first, the finite system-environment leads to a modified environment state before the measurement and establishes correlations between the system and the environment, and second, because of these correlations, the projective measurement on the system affects the environment state. It should be noted that such an initial system-environment state has been considered previously \cite{PazPRA1997, ErezNature2008, GordonNJP2009, GordonNJP2010, MorozovPRA2012, ChaudhryPRA2013}. This initial state then evolves under the action of the total Hamiltonian $H$, re-establishing correlations between the system and the environment, and the total equilibrium state $e^{-\beta H}/Z$ is eventually obtained.

Let us now investigate the initial state in more detail. We can perform a perturbative expansion of the initial state given by Eq.~\eqref{withcorrinitialstate} in powers of $\alpha$. To this end, we invoke the Kubo identity, which states that, given two operators $X$ and $Y$,
\begin{equation}
e^{\beta(X + Y)} = e^{\beta X} \left( 1 + \int_0^\beta d\lambda e^{-\lambda X} Y e^{\lambda(X + Y)} \right).
\end{equation}
By setting $X = -(H_S + H_B)$ and $Y = -V$, we obtain, to first order in the system-environment coupling strength,
\begin{align}
e^{-\beta H} &\approx e^{-\beta(H_S + H_B)} \times \notag \\
&\left[ 1 - \int_0^\beta d\lambda e^{\lambda(H_S + H_B)}\alpha V e^{-\lambda (H_S + H_B)}\right].
\end{align}
Assuming that $V$ can be written in the form $V = F \otimes B$, where $F$ ($B$) is an operator acting in the system (bath) Hilbert space \cite{footnoteonV},
\begin{align}
&\opav{\psi}{e^{-\beta H}}{\psi} = \opav{\psi}{e^{-\beta H_S}}{\psi} e^{-\beta H_B} - \alpha e^{-\beta H_B} \notag \\
&\times \int_0^\beta d\lambda e^{\lambda H_B}Be^{-\lambda H_B} \opav{\psi}{e^{-\beta H_S} e^{\lambda H_S} F e^{-\lambda H_S}}{\psi}.
\end{align}
We write this as
\begin{equation}
\label{initialenvstate}
\opav{\psi}{e^{-\beta H}}{\psi} = \opav{\psi}{e^{-\beta H_S}}{\psi} e^{-\beta H_B} - \alpha e^{-\beta H_B} E(\beta),
\end{equation}
where $E(\beta)$ is an operator acting in the Hilbert space of the environment only. Physically, $E(\beta)$ is essentially the first order change in the environment state as a result of the initial correlations. This modification can be zero; for example, as will see in more detail in the next section, for the model considered in Refs.~\cite{ErezNature2008, GordonNJP2009, GordonNJP2010}, the first order modification to the environment is zero. It should also be noted that the initial environment state [$\sim \opav{\psi}{e^{-\beta H}}{\psi}$]
is not the reduced state of the total system-environment equilibrium state [$\sim \text{Tr}_S[e^{-\beta H}]$]. As a result, after the system state preparation, the environment evolves and approaches this equilibrium environment state. This evolution of the environment can have an important dynamical consequence for the system dynamics, as we shall see below.

Before proceeding, we write for convenience
\begin{equation}
\rho_{\text{tot}}(0) = \rho(0) \otimes \left[ \rho_B^{(0)} + \rho_B^{(1)} + \hdots \right],
\end{equation}
with the superscript denoting the order of the coupling strength. It should be noted that as $\beta \rightarrow 0$ (that is, we approach high temperatures), $E(\beta) \rightarrow 0$. This is what we intuitively expect - at high temperatures, the effect of initial correlations becomes less and less significant.

With these preliminary calculations out of the way, we now proceed to the main task of deriving the master equation. The first part is easy [see mainly Eq.~(\ref{allterms})] - this is simply
\begin{align}
&\text{Tr}_{\text{S,B}} \left[ \rho_{\text{tot}}(0) i [H_0^H(t),X_{nm}^H(t)] \right] \notag \\
&= i \text{Tr}_{\text{S,B}} \left[ U(t) \rho_{\text{tot}}(0) U^\dagger(t) [H_0,X_{nm}]\right] \notag \\
&= i \text{Tr}_S \left[ \rho(t) (H_S \ket{n}\bra{m} - \ket{n}\bra{m}H_S)\right] \notag \\
&= i \opav{m}{[\rho(t),H_S]}{n}.
\end{align}
Physically, this term represents the evolution of the system due to the uncoupled system Hamiltonian $H_S$.
Moving on, the next term [arising from the second term on the right hand side of Eq.~(\ref{allterms})]
is
$$ i\alpha \text{Tr}_{\text{S,B}} \left[ \rho_{\text{tot}}(0) U_0^\dagger(t) [V, X_{nm}]U_0(t) \right].$$
To second order in the coupling strength, only $\rho_B^{(0)}$ and $\rho_B^{(1)}$ contribute to the master equation. The contribution of $\rho_B^{(0)}$ is
\begin{align}
&i\alpha \text{Tr}_{\text{S,B}} \left[ (\rho(0) \otimes \rho_B^{(0)}) U_0^\dagger(t) [F \otimes B, X_{nm}] U_0(t) \right] \notag \\
&= i\alpha \text{Tr}_{\text{S,B}} \left[ (\rho(0) \otimes \rho_B^{(0)})U_0^\dagger(t)([F,X_{nm}]\otimes B)U_0(t) \right] \notag.
\end{align}
The trace over the bath gives a term proportional to $\langle U_B^\dagger(t) B U_B(t) \rangle_B$, where $\langle \hdots \rangle_B$ denotes an average taken with respect to the bath state $\rho_B = e^{-\beta H_B}/Z_B$. This is usually zero. Even if it is not zero, the contribution of this term can be absorbed into the system Hamiltonian.
We now note that
\begin{align}
Z &= Z_B \opav{\psi}{e^{-\beta H_S}}{\psi} - \alpha Z_B \langle E(\beta)\rangle_B \notag \\
&= Z_B Z',
\end{align}
where $Z' = \opav{\psi}{e^{-\beta H_S}}{\psi} - \alpha \langle E(\beta)\rangle_B$ is the modification of the partition function due to the finite coupling strength and the projective measurement. It should be noted that, just like $\langle U_B^\dagger(t) B U_B(t) \rangle_B$, $\langle E(\beta) \rangle_B$ is generally zero.
It follows from Eq.~\eqref{initialenvstate} that
\begin{equation}
\label{envstatefirstordercorrection}
\rho_B^{(1)} = -\frac{1}{Z_BZ'} \alpha e^{-\beta H_B} E(\beta).
\end{equation}
Using this expression of $\rho_B^{(1)}$ and returning to the
second term of the right-hand side of Eq.~(9), we get
\begin{align}
&-\frac{i\alpha^2}{Z_B Z'} \text{Tr}_S \left[ \rho(0) U_S^\dagger(t) [F, \ket{n}\bra{m}] U_S(t) \right] \notag \\
&\times \text{Tr}_B \left[ e^{-\beta H_B} E(\beta) U_B^\dagger (t) B U_B(t) \right] \notag \\
&= -\frac{i\alpha^2}{Z'} \langle E(\beta) \widetilde{B}(t)\rangle_B \text{Tr}_S \left[ \rho(t) [F,\ket{n}\bra{m}]\right] \notag \\
&= -\frac{i\alpha^2}{Z'} \langle E(\beta) \widetilde{B}(t)\rangle_B \opav{m}{[\rho(t),F]}{n},
\end{align}
where $U_S(t) \rho(0) U_S^\dagger(t)$ has been replaced by $\rho(t)$ since the corrections give us terms of higher order in the coupling strength. In physical terms, this terms arises because the modified environment state evolves back to the equilibrium environment state. However, since the environment and the system are coupled, this evolution of the environment also affects the system evolution.

For the next term, to second order, only $\rho_B^{(0)}$ contributes. But then $Z'$ is simply $\opav{\psi}{e^{-\beta H_S}}{\psi}$, so $\rho_B^{(0)} = \rho_B$. Therefore, we obtain the same term as in the standard second order master equation, the derivation of which can be found in Appendix \ref{nocorrmasterequationappendix}. Compared with the standard master equation, the term that we need to concentrate is then (we now set $\alpha = 1$)
\begin{equation}
f_{\text{corr}}(t) = \frac{\langle E(\beta) \widetilde{B}(t)\rangle_B}{Z'}.
\end{equation}
The complete master equation can then be written as
\begin{align}
\label{masterequationwithcorr}
\frac{d\rho(t)}{dt} &= i [\rho(t),H_S] - if_{\text{corr}}(t) [\rho(t),F] \, + \notag \\
&\int_{0}^t ds \lbrace [\bar{F}(t,s)\rho(t),F]C_{ts} + h.c. \rbrace,
\end{align}
where,
\begin{align}
\bar{F}(t,s) &= U_{\text{S}}(t,s)FU_{\text{S}}^\dagger(t,s), \\
C_{ts} &= \langle \widetilde{B}(t) \widetilde{B}(s) \rangle_B, \\
\widetilde{B}(t) &= U_{\text{B}}^\dagger(t)  B U_{\text{B}}(t).
\end{align}
The structure of this master equation leads to the hermiticity and trace of $\rho$ being preserved (see also Appendix \ref{proofoffcorrreal}). It is interesting to note that the contribution due to the initial correlations is of the same structure as that of a coherent term. However, for an unknown bath (which is true in most cases), this term induced by a system-state preparation is still
undesired because it is normally unknown beforehand.

In contrast, the master equation obtained if we start from the uncorrelated initial state given by \eqref{nocorrinitialstate} would be
\begin{align}
\label{masterequationnocorr}
\frac{d\rho(t)}{dt} &= i [\rho(t),H_S] \, + \notag \\
&\int_{0}^t ds \lbrace [\bar{F}(t,s)\rho(t),F]C_{ts} + h.c. \rbrace.
\end{align}
It should be noted that at no point have we made any assumption regarding the memory of the environment. We again emphasize that this is a second-order master equation in terms of the system-bath coupling strength.

Before moving on, let us use our physical intuition to guess when the effect of $f_{\text{corr}}(t)$ will be significant. Three key conditions need to be satisfied: first, the post-measurement environment state needs to be considerably different from the thermal bath state; second, the environment should have a long correlation time, that is, it should be non-Markovian, so that it does not forget too quickly what its initial state was; and third, the initial state preparation should be such that $[\rho(0), F] \neq 0$, otherwise, it could be that the initial correlations do not get the chance to play a significant role in the system dynamics. In what follows, we will show that when these conditions are satisfied, the initial correlations can indeed play an important role.

\section{Application to a large spin model}

We now apply our master equation to study the system dynamics in a variant of the paradigmatic spin-boson model \cite{Weissbook} extended to many spins interacting with a common environment. The total system-environment Hamiltonian can be written as
\begin{equation}
\label{modelhamiltonian}
H = H_S + H_B + V,
\end{equation}
with
\begin{align}
H_S = \varepsilon J_z + \Delta J_x, \, H_B = \sum_k \omega_k b_k^\dagger b_k, \\
V = J_x \sum_k (g_k^*b_k + g_k b_k^\dagger).
\end{align}
Here the $J_{x,y,z}$ operators are collective spin operators with $J_x^2 + J_y^2 + J_z^2 = \frac{N}{2}(\frac{N}{2} + 1)$, $\omega_0$ is the energy bias, $\Delta$ is the tunneling amplitude, $H_B$ describes a bath of harmonic oscillators (ignoring the zero-point energy), while $V$ describes the interaction between the spin system and the common harmonic oscillator bath. Such large spin Hamiltonians have been used, for instance, in the study of a two-mode BEC interacting via collisions with thermal atoms \cite{KurizkiPRL2011}, as well as in modeling the dynamics of intrinsic spins and to describe transport in double quantum dot arrays \cite{VorrathChemPhys2004, VorrathPRL2005}.  We set $\hbar=1$ throughout and the values of other parameters will be in dimensionless units.

We choose our initial state $\ket{\psi}$ to be such that $J_z \ket{\psi} = -\frac{N}{2}\ket{\psi}$. We denote such a state as $\ket{-N/2}$. This state is chosen for three reasons. First, it is relatively simple to prepare experimentally - all the two-level atoms are in the same initial state. Second, the computation of $f_{\text{corr}}(t)$ is not too complicated. Third, and most importantly, since each atom is in the same initial state, we expect that each atom affects the environment in the same way. Therefore, even though each individual atom may be coupled to the environment weakly, the collective environment as a whole may be substantially different from the thermal state, depending on the number of two-level atoms.

To begin, we note that $F = J_x$, and $B = \sum_k (g_k^*b_k + g_k b_k^\dagger)$. Our task is to evaluate $E(\beta)$. In order to do so, we first evaluate
\begin{equation}
e^{\lambda H_S} J_x e^{-\lambda H_S} = a_x J_x + a_y J_y + a_z J_z,
\end{equation}
where
\begin{align}
a_x = \frac{\Delta^2 + \varepsilon^2 \cosh(\lambda \widetilde{\Delta})}{\widetilde{\Delta}^2}, \\
a_y = \frac{i\varepsilon}{\widetilde{\Delta}} \sinh(\lambda \widetilde{\Delta}), \\
a_z = \frac{\varepsilon\Delta}{\widetilde{\Delta}^2} [1 - \cosh(\lambda \widetilde{\Delta})],
\end{align}
with $\widetilde{\Delta} \equiv \sqrt{\Delta^2 + \varepsilon^2}$.
To calculate the inner product (that is, $\opav{\psi}{e^{-\beta H_S}}{\psi}$), it is useful to write \cite{puribook}
\begin{equation}
e^{-\beta H_S} = e^{f J_+} e^{f_z J_z} e^{f J_-},
\end{equation}
where
\begin{align}
f = - \frac{\Delta}{\widetilde{\Delta}} \frac{\sinh(\beta \widetilde{\Delta}/2)}{\mu},
\end{align}
with
\begin{equation}
\mu = \cosh(\beta \widetilde{\Delta}/2) + \frac{\varepsilon}{\widetilde{\Delta}} \sinh(\beta \widetilde{\Delta}/2),
\end{equation}
and
\begin{equation}
f_z = -2 \ln \mu.
\end{equation}
Using the properties of the raising and lowering angular momentum operators, it can be shown that
\begin{align}
&\opav{\psi}{e^{-\beta H_S} e^{\lambda H_S} F e^{-\lambda H_S}}{\psi} = \notag \\
&\mu^{N - 1} \frac{N}{2} \left[ \kappa + \frac{\varepsilon \Delta}{\widetilde{\Delta}^2} \cosh(\lambda \widetilde{\Delta} - \beta \widetilde{\Delta}/2)\right],
\end{align}
where
\begin{equation}
\kappa \equiv -\frac{\Delta}{\widetilde{\Delta}} \sinh\left(\frac{\beta \widetilde{\Delta}}{2}\right) - \frac{\varepsilon \Delta}{\widetilde{\Delta}^2} \cosh\left(\frac{\beta \widetilde{\Delta}}{2}\right).
\end{equation}
We also have that
\begin{equation}
Z' = \opav{\psi}{e^{-\beta H_S}}{\psi} = \mu^N.
\end{equation}
It follows that
\begin{align}
&\frac{\opav{\psi}{e^{-\beta H_S} e^{\lambda H_S} F e^{-\lambda H_S}}{\psi}}{Z'} = \notag \\
& \frac{N}{2} \left[ A + \mathcal{B} \cosh(\lambda \widetilde{\Delta} - \mathcal{C})\right],
\end{align}
where
\begin{equation}
A \equiv \frac{\kappa}{\mu}, \, \mathcal{B} = \frac{\varepsilon \Delta}{\mu \widetilde{\Delta}^2}, \, \mathcal{C} = \beta \widetilde{\Delta}/2.
\end{equation}
We also know that
\begin{equation}
e^{\lambda H_B} B e^{-\lambda H_B} = \sum_k (g_k^* b_k e^{-\lambda \omega_k} + g_k b_k^\dagger e^{\lambda \omega_k}).
\end{equation}
We can then write
\begin{equation}
\frac{E(\beta)}{Z'} = \frac{N}{2} \sum_k [g_k^* b_k Q_1(\beta,\omega_k) + g_k b_k^\dagger Q_2(\beta,\omega_k)],
\end{equation}
with
\begin{align}
&Q_1(\beta,\omega_k) = \int_0^\beta d\lambda e^{-\lambda \omega_k}[A + \mathcal{B} \cosh(\lambda \widetilde{\Delta} - \mathcal{C})],\\
&Q_2(\beta,\omega_k) = \int_0^\beta d\lambda e^{\lambda \omega_k} [A + \mathcal{B} \cosh(\lambda \widetilde{\Delta} - \mathcal{C})].
\end{align}
Using the fact that $[b_k,b_{k'}^\dagger] = \delta_{kk'}$, $[b_k,b_{k'}] = [b_k^\dagger,b_{k'}^\dagger] = 0$  and $ \langle b_k^\dagger b_k \rangle_B = \frac{1}{e^{\beta \omega_k} - 1} \equiv n_k$, we find that
\begin{align}
&f_{\text{corr}}(t) = \frac{N}{2} \sum_k |g_k|^2 \, \times \notag \\
&\left[ Q_1(\beta,\omega_k)e^{i\omega_k t}(1 + n_k) + Q_2(\beta,\omega_k) e^{-i\omega_k t}n_k \right].
\end{align}
We can then show that the imaginary part is zero as expected, and, after evaluating the real part, we can finally write
\begin{align}
&f_{\text{corr}}(t) = N \sum_k |g_k|^2 \cos(\omega_k t) \times \notag \\
& \left\lbrace \frac{A}{\omega_k} + \frac{D}{\widetilde{\Delta}^2 - \omega_k^2}\left[\widetilde{\Delta} \coth\left(\frac{\beta \omega_k}{2}\right) - \omega_k \coth\left(\frac{\beta\widetilde{\Delta}}{2}\right) \right]\right\rbrace,
\end{align}
with
\begin{align}
A = -\frac{\Delta}{\widetilde{\Delta}} \frac{\widetilde{\Delta} + \varepsilon \coth(\beta \widetilde{\Delta}/2)}{\widetilde{\Delta} \coth(\beta\widetilde{\Delta}/2) + \varepsilon}, \\
D = \frac{\varepsilon\Delta/\widetilde{\Delta}}{\widetilde{\Delta} \coth(\beta\widetilde{\Delta}/2) + \varepsilon}.
\end{align}
Let us emphasize the physical origin of this term in the master equation. The equilibrium bath state is modified as a result of the initial correlations and the projective measurement. Since the oscillators are no longer in equilibrium, they start to evolve. However, since these oscillators are coupled to the two-level atoms, this dynamical evolution of the oscillators affects the evolution of the atoms as well. It is precisely this evolution of the atoms that is captured by $f_{\text{corr}}(t)$. It is important to note that $f_{\text{corr}}(t)$ derived above is proportional to $N$. That is,
the number of two-level systems in the ensemble amplifies the system-bath correlation effect.
In addition, it is a simple exercise in algebra to show that, as expected, this expression for $f_{\text{corr}}$ tends to zero as $\beta \rightarrow 0$.

\begin{figure}[b]
   \includegraphics[scale = 1]{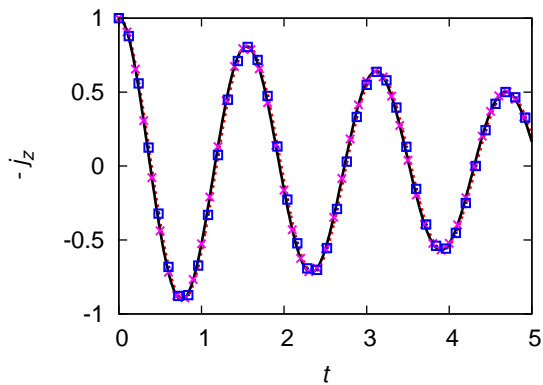}
   \centering
  	\caption{(color online) Behavior of $-j_z$ versus $t$ for $N = 1$ using the exact solution with (magenta crosses) and without (blue squares) initial correlations, as well as using the master equation with (solid, black line) and without (dotted, red line) initial correlations. We have used $\Delta = 4$, $G = 0.05$ and $\omega_c = 5$. Here and in all other figures, the plotted variables are all in dimensionless units.}
  	\label{dephasingJ0p5}
\end{figure}

\begin{figure}[t]
   \includegraphics[scale = 1]{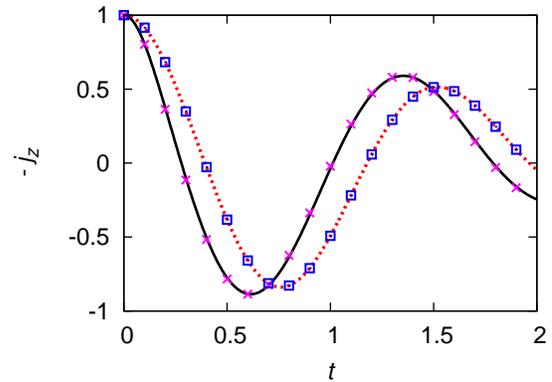}
   \centering
  	\caption{(color online) Same as Fig.~\ref{dephasingJ0p5}, except that we now have $N = 10$.}
  	\label{dephasingJ5}
\end{figure}

Before proceeding, it is useful to look at two limiting cases: \\
i) Dicke model \cite{Dicke1954}. In this case, $\Delta = 0$. We then get $A = \mathcal{B} = 0$, whereby $E(\beta) = 0$. Therefore, the initial correlations have no effect in this case because the environment state is the same as the thermal bath state. It should be noted that for $N = 1$, the Dicke model is the same as the model used in Refs.~\cite{ErezNature2008, GordonNJP2009, GordonNJP2010}, and thus the use of the master equation \eqref{masterequationnocorr} therein is justified.\\
ii) Pure dephasing model. In this case, $\varepsilon = 0$. We then find that $D = 0$, while $A = -\tanh(\beta \Delta/2)$. This model can also be solved exactly [see Appendix \ref{puredephasingappendix}], thereby serving as a useful benchmark for our master equation. Therefore, before turning to more general cases, we compare the performance of our master equation against the exact solution. As usual, we replace the sum over the different modes $k$ by an integration, that is, we make the substitution $\sum_k |g_k|^2 C(\omega_k) \rightarrow \int_0^\infty J(\omega) C(\omega)$, where $J(\omega)$ is the spectral density of the environment. We choose the spectral density to be Ohmic with exponential cutoff, namely $J(\omega) = G\omega e^{-\omega/\omega_c}$. Throughout this paper, we set $\beta = 1$.

\begin{figure}[t]
   \includegraphics[scale = 1]{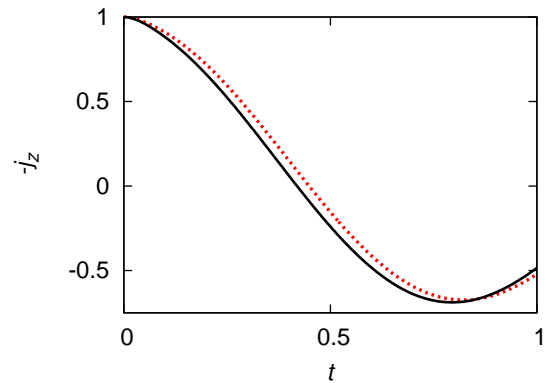}
   \centering
  	\caption{(color online) Behavior of $-j_z$ against $t$ for $N = 2$ with (black, solid) and without (dotted, red) taking into account initial correlations. Here we have used $\Delta = 3.5$ and $\varepsilon = 0.5$, while the rest of the parameters used are the same as those in Fig.~\ref{dephasingJ0p5}.}
  	\label{fullmodelepsilon0p5J1}
\end{figure}

\begin{figure}[t]
   \includegraphics[scale = 1]{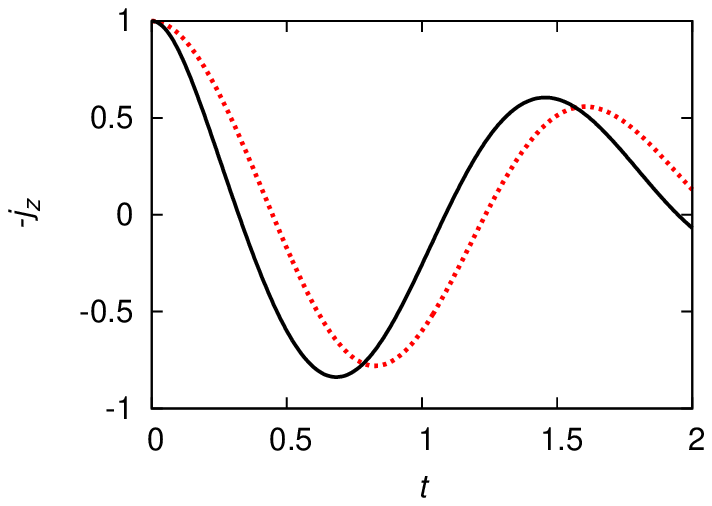}
   \centering
  	\caption{(color online) Same as Fig.~\ref{fullmodelepsilon0p5J1}, except that we now have $N = 10$.}
  	\label{fullmodelepsilon0p5J5}
\end{figure}

\begin{figure}[t]
   \includegraphics[scale = 1]{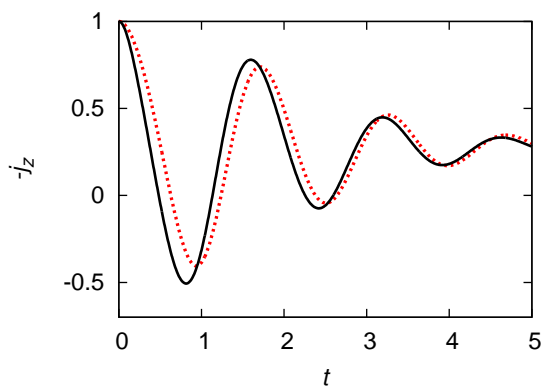}
   \centering
  	\caption{(color online) Same as Fig.~\ref{fullmodelepsilon0p5J5}, except that we now have $\varepsilon = 1.5$ and $\Delta = 2.5$.}
  	\label{fullmodelepsilon1p5J5}
\end{figure}

\begin{figure}[t]
   \includegraphics[scale = 1]{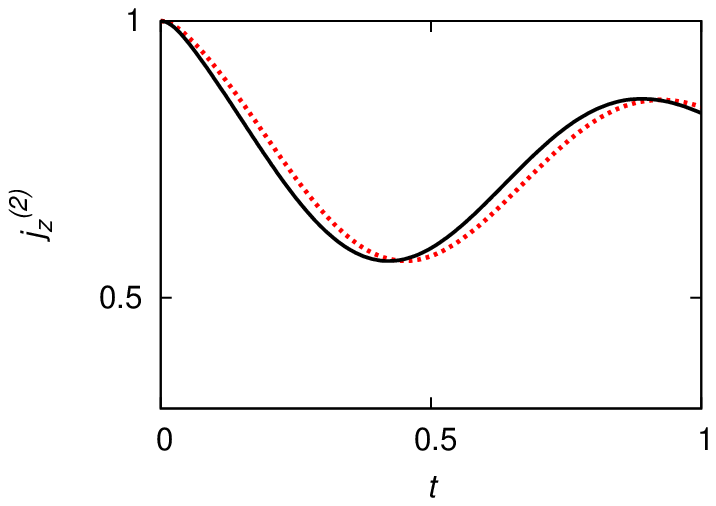}
   \centering
  	\caption{(color online) Behavior of $j_z^{(2)}$ against $t$ for $N = 2$ with (black, solid) and without (dotted, red) taking into account initial correlations. The rest of the parameters used are the same as those in Fig.~\ref{fullmodelepsilon0p5J1}.}
  	\label{jzsqJ1}
\end{figure}

\begin{figure}[t]
   \includegraphics[scale = 1]{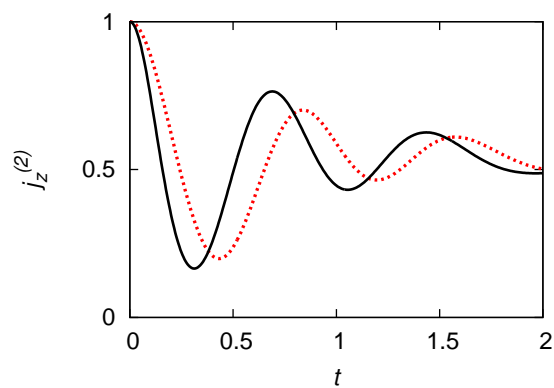}
   \centering
  	\caption{(color online) Same as Fig.~\ref{jzsqJ1}, except that we now have $N = 10$.}
  	\label{jzsqJ5}
\end{figure}

In Fig.~\ref{dephasingJ0p5}, we have plotted the behavior of $-j_z \equiv -2\langle J_z \rangle /N$ against $t$ using the master equation with and without taking into account initial correlations, as well as the dynamics obtained using the exact solution for $N = 1$. As can be seen, in this case, the initial correlations play an insignificant role - the dynamics, both from the master equation and using the exact solution, are the same for all intents and purposes. Essentially, the reason is that with only one atom interacting weakly with the environment, the environment is hardly modified, and thus $f_{\text{corr}}(t)$ plays a negligible role. However, as illustrated in Fig.~\ref{dephasingJ5}, with increasing $N$, the effect of initial correlations becomes more significant, as we expected. There is now a significant difference between the dynamics with and without initial correlations, signifying that initial correlations now play an important role. Moreover, the  master equation is able to reproduce the exact dynamics, both with and without initial correlations.

Now that we are confident that in the weak coupling regime, our master equation is able to capture well the effect of initial correlations, we move beyond the exactly solvable dephasing model. More specifically, we now consider a finite $\varepsilon$. As shown in Fig.~\ref{fullmodelepsilon0p5J1}, for a finite value of $\varepsilon$ with $N = 2$, there is a small effect of the initial correlations, again because for a small $N$, the state of the environment is hardly affected. However, it is illustrated in Figs.~\ref{fullmodelepsilon0p5J5} and \ref{fullmodelepsilon1p5J5} that by increasing $N$ to $N = 10$ the effect of the initial-state preparation becomes more pronounced, even for a non-zero value of $\varepsilon$. We expect that the effect of initial correlations increases still further as we increase $N$. Moreover, the influence of initial correlations also slowly decreases as we increase $\varepsilon$ such that, as we argued previously, in the Dicke model limit, the initial correlations do not play any role in the system dynamics. It should also be noted that, as shown in Fig.~\ref{fullmodelepsilon1p5J5}, the effect of initial correlations at longer times becomes smaller and smaller. This makes sense physically since, after some time has passed, the system should forget its initial state. Finally, to show that the effect of initial correlations are not manifested in the dynamics of $j_z$ alone, we show in Figs.~\ref{jzsqJ1} and \ref{jzsqJ5} the dynamics of $j_z^{(2)} \equiv 4\langle J_z^2 \rangle/N^2$. Such an observable is relevant in the study of spin squeezing and entanglement (see, for example, Ref.~\cite{TothPRA2009}). Once again, the initial correlations have a noticeable effect on the dynamics, an effect that increases with increasing $N$.
Note however, numerically speaking, the calculations for an even larger $N$ at long times would be demanding. Luckily, we can use our master equation to obtain the system dynamics at short times approximately, thereby showing the effect of initial correlations for even larger $N$.

\section{Short time approximation}

\begin{figure}[h]
   \includegraphics[scale = 1]{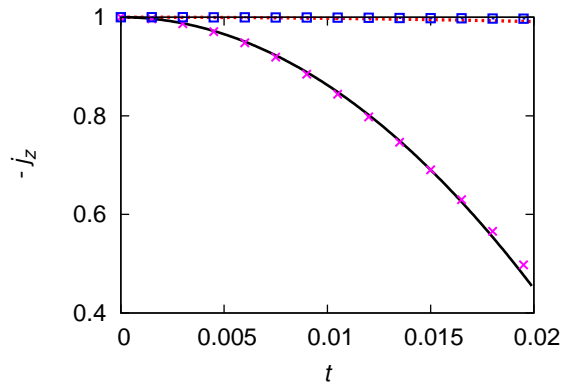}
   \centering
  	\caption{(color online) Behavior of $-j_z$ versus $t$ for $N = 1000$ using the exact solution with (magenta crosses) and without (blue squares) initial correlations, as well as using the short time approximation with (solid, black line) and without (dotted, red line) initial correlations. We have used $\Delta = 4$, $\varepsilon = 0$, $G = 0.05$ and $\omega_c = 5$.}
  	\label{smalltimeapproximationpuredephasing}
\end{figure}

\begin{figure}[h]
   \includegraphics[scale = 1]{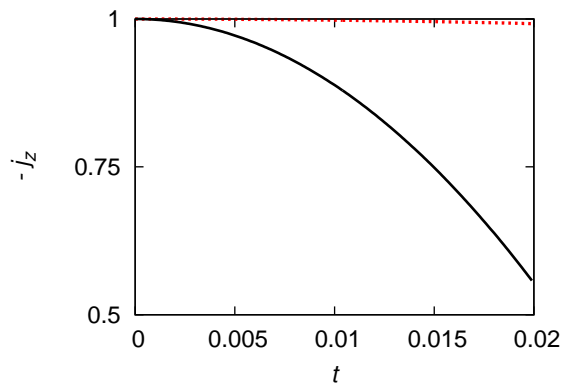}
   \centering
  	\caption{(color online) Dynamics of $-j_z(t)$ with (solid, black) and without (dotted, red) initial correlations for short times. Here we have used $\varepsilon = 0.5$ and $\Delta = 3.5$. The rest of the parameters are the same as Fig.~\ref{smalltimeapproximationpuredephasing}.}
  	\label{smalltimeapproximationepsilon0p5}
\end{figure}

To gain deeper understanding of the system dynamics due to the initial system-environment correlations, we investigate in more detail the system dynamics at short times \cite{ErezNature2008}. To do this, we first assume that $H_S$ does not depend explicitly on time and write the master equation \eqref{masterequationwithcorr} as
\begin{align}
\frac{d\rho(t)}{dt} &= i [\rho(t),H_S'(t)] \, + \notag \\
&\int_{0}^t d\tau \lbrace [\bar{F}(\tau)\rho(t),F]C(\tau) + h.c. \rbrace,
\end{align}
where $\tau = t - s$ and $H_S'(t) = H_S - f_{\text{corr}}(t)F$. It follows that for small time $t$,
\begin{align}
\label{smalltimeapproxgeneral}
&\rho(t) \approx \rho + i[\rho, H_S']t \, + \notag \\
&\frac{t^2}{2} \left\lbrace [H_S',[\rho, H_S']] + C(0)[2F\rho F - F^2\rho - \rho F^2] \right\rbrace,
\end{align}
where $\rho = \rho(0)$ for brevity and $H_S' = H_S'(0)$. An equivalent expression for the evolution of the density matrix at short times after neglecting initial correlations is found by setting $f_{\text{corr}}(0) = 0$. Considering the initial state $\rho(0) = \ket{-N/2}\bra{-N/2}$, we find that
\begin{equation}
-j_z(t) \approx 1 - \frac{t^2}{2} \left[C(0) + \Delta^2 + f_{\text{corr}}^2(0) - 2\Delta f_{\text{corr}}(0) \right].
\end{equation}
The important point here is that $f_\text{corr}(0)$ increases as $N$ increases, while $C(0)$ and $\Delta$ obviously do not. The system evolution is then dominated by $f_{\text{corr}}(0)$ for large $N$ (note that $f_{\text{corr}}(0)$ is negative for our choice of initial state). Moreover, $f_{\text{corr}}(0)$ depends on the details of the environment, which generally we do not know. On the other hand, without initial correlations, we have
\begin{equation}
-j_z(t) \approx 1 - \frac{t^2}{2} \left[C(0) + \Delta^2 \right].
\end{equation}
We now use the exactly solvable dephasing model to check the validity of our small-time approximation. As shown in Fig.~\ref{smalltimeapproximationpuredephasing}, we are able to reproduce the system dynamics at small times very well. We then use this approach to find the system dynamics for large $N$ at small times in the non-exactly solvable case [see Fig.~\ref{smalltimeapproximationepsilon0p5}].  It is seen that with a large $N$,
the dynamics accounting for the nonequilibrium state of the environment due to system-state preparation is drastically different from the case without accounting for system-bath correlation.

Finally, we examine what this rapid change of state implies for quantum control \cite{LloydDD, DavidsonPRA2009, largespinmitchell}. One of the central objectives of control is to apply control fields in order to preserve the quantum state. As we have seen, for large $N$, the quantum state can evolve rapidly in an unknown manner if the bath properties are not available. It then becomes imperative that the control fields are applied with a shorter time-scale in mind. In particular, we have seen that, starting from $j_z = -1$, the change in $j_z$ increases as $N$ increases at short times. This means that, in order to keep $j_z$ close to $-1$, we need to apply control fields on a time-scale that reduces as $N$ increases. More specifically, if we are using pulses, the first pulse should be applied within a time scale such that $\frac{t^2}{2}[C(0) + \Delta^2 + f_{\text{corr}}^2(0) - 2\Delta f_{\text{corr}}(0)]$ is as small as possible. Because $f_{\text{corr}}(0)$ is proportional to $N$, the short time scale needed can be challenging experimentally for a large $N$. It should be noted, however, that this result depends on the choice of initial state and the system-environment model. As we will explicitly show in the next section, the effect of initial correlations could be still small sometimes.

It is also important to note in passing that, unlike Ref.~\cite{ErezNature2008}, the contribution of the first order term in Eq.~\eqref{smalltimeapproxgeneral} is not zero. For example, for the observable $j_y \equiv 2\langle J_y \rangle/N$, we obtain for the initial state $\ket{-N/2}\bra{-N/2}$,
\begin{equation}
j_y(t) \approx [\Delta - f_{\text{corr}}(0)]t.
\end{equation}

\section{Different state preparation}

\subsection{$\ket{\psi} = \ket{N/2}$}
In this case the initial state is polarized in the opposite direction.
We once again need to calculate the effect of the initial correlations. The calculation is almost the same as before, but there are a few notable differences. Here we only present the final result, and defer the details to Appendix \ref{appendixcalculations}. We find that with the initial state $\ket{\psi} = \ket{N/2}$, we have
\begin{align}
&f_{\text{corr}}(t) = N \sum_k |g_k|^2 \cos(\omega_k t) \times \notag \\
& \left\lbrace \frac{A}{\omega_k} + \frac{D}{\widetilde{\Delta}^2 - \omega_k^2}\left[\widetilde{\Delta} \coth\left(\frac{\beta \omega_k}{2}\right) - \omega_k \coth\left(\frac{\beta\widetilde{\Delta}}{2}\right) \right]\right\rbrace,
\end{align}
with
\begin{align}
A = -\frac{\Delta}{\widetilde{\Delta}} \frac{\widetilde{\Delta} - \varepsilon \coth(\beta \widetilde{\Delta}/2)}{\widetilde{\Delta} \coth(\beta\widetilde{\Delta}/2) - \varepsilon}, \\
D = - \frac{\varepsilon\Delta/\widetilde{\Delta}}{\widetilde{\Delta} \coth(\beta\widetilde{\Delta}/2) - \varepsilon}.
\end{align}
These results should be compared with the ones obtained before. Specifically, it should be noted that $f_{\text{corr}}(t)$, for the same values of $\varepsilon$ and $\Delta$, is different for the cases $\ket{\psi} = \ket{-N/2}$ and $\ket{\psi} = \ket{N/2}$. The contribution of the initial correlations in the master equation itself changes depending on the initial state preparation, which means that the effect of initial correlations depends on the initial state preparation.

\begin{figure}[h]
   \includegraphics[scale = 1]{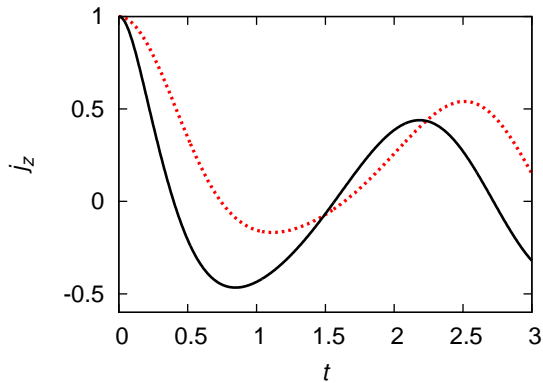}
   \centering
  	\caption{(color online) Graph of $j_z$ against $t$, starting from the state $\ket{\psi} = \ket{N/2}$. The parameters used are the same as in Fig.~\ref{fullmodelepsilon1p5J5}.}
  	\label{upstatejz}
\end{figure}

In Fig.~\ref{upstatejz}, we plotted $j_z$ against time, starting from the state $\ket{\psi} = \ket{N/2}$. Once again, it is seen that the initial correlations play a significant role in the dynamics. It is instructive to note the asymmetry between Fig.~\ref{fullmodelepsilon1p5J5} and Fig.~\ref{upstatejz}. This asymmetry is due to two reasons. First, the coherent evolution (that is, the evolution due to $H_S$ alone) itself causes asymmetry in the dynamics of $j_z$, a fact that is easily visualized in the Bloch vector picture. Secondly, as we have noted before, the influence of initial correlations is different for the two cases.

\subsection{Each spin prepared in a coherent superposition}
We now consider a different state preparation, namely $\ket{\psi}$ such that $J_x \ket{\psi} = \frac{N}{2} \ket{\psi}$. This is a clearly an eigenstate of $J_x$ with eigenvalue $\frac{N}{2}$. In order to perform the calculation for the effect of the initial correlations, it is useful to first rotate our axes so that we now have, in the rotated frame,
\begin{equation}
H^R = H^R_S + H_B + V^R,
\end{equation}
with
\begin{align}
H^R_S = \varepsilon_r J_z + \Delta_r J_x, \, H_B = \sum_k \omega_k b_k^\dagger b_k, \\
V^R = J_z \sum_k (g_k^*b_k + g_k b_k^\dagger),
\end{align}
where $\varepsilon_r = \Delta$ and $\Delta_r = -\varepsilon$, and our initial state is now $\ket{\psi^R}$, which is an eigenstate of $J_z$ with eigenvalue $\frac{N}{2}$. It can then be shown that
\begin{align}
&f_{\text{corr}}(t) = N \sum_k |g_k|^2 \cos(\omega_k t) \times \notag \\
& \left\lbrace \frac{A}{\omega_k} + \frac{D}{\widetilde{\Delta}_r^2 - \omega_k^2}\left[\widetilde{\Delta}_r \coth\left(\frac{\beta \omega_k}{2}\right) - \omega_k \coth\left(\frac{\beta\widetilde{\Delta}_r}{2}\right) \right]\right\rbrace,
\end{align}
with
\begin{align}
A = -\frac{\varepsilon_r}{\widetilde{\Delta}_r} \frac{\widetilde{\Delta}_r - \varepsilon_r \coth(\beta \widetilde{\Delta}_r/2)}{\widetilde{\Delta}_r \coth(\beta\widetilde{\Delta}_r/2) - \varepsilon_r}, \\
D = \frac{\Delta_r^2}{\widetilde{\Delta}_r^2} \frac{1}{\coth(\beta\widetilde{\Delta}_r/2) - \varepsilon_r/\widetilde{\Delta}_r},
\end{align}
and $\widetilde{\Delta}_r \equiv \sqrt{\varepsilon_r^2 + \Delta_r^2}$ (see Appendix \ref{appendixcalculations} for details).

\begin{figure}[h]
   \includegraphics[scale = 1]{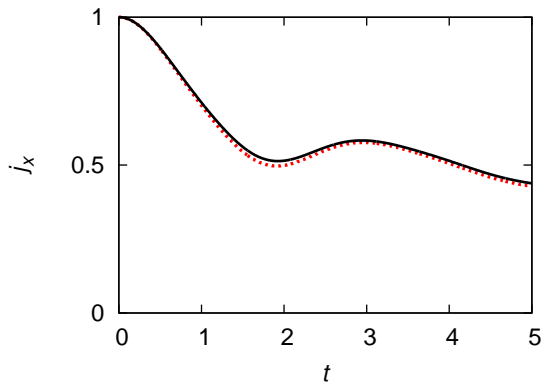}
   \centering
  	\caption{(color online) $j_x$ versus $t$ with (solid, black) and without (dotted, red) taking into account initial correlations starting from state $\ket{\psi}$ such that $J_x\ket{\psi} = \frac{N}{2}\ket{\psi}$. Here we have used $N = 10$, $\Delta = 3$, $\varepsilon = 1$, $G = 0.05$ and $\omega_c = 5$.}
  	\label{statesalignedjx}
\end{figure}

In Fig.~\ref{statesalignedjx} we show the dynamics with and without initial correlations. This time we find that the initial correlations play an insignificant role. This is not simply due to the factor $f_{\text{corr}}(t)$ being small - this factor is certainly significant at short times. However, if we look at the master equation closely, we notice that the effect of the initial correlations is incorporated via the term
$$ - if_{\text{corr}}(t)[\rho(t), F]. $$
For the initial state preparation $\ket{\psi}$ such that $J_x \ket{\psi} = \frac{N}{2} \ket{\psi}$, with the system-environment Hamiltonian given by Eq.~\eqref{modelhamiltonian}, we find that $[\rho(0), F] = 0$. Therefore, at time $t = 0$, the effect of initial correlations is zero. By the time the state evolves to a state $\rho(t)$ such that $[\rho(t), F]$ is appreciably different from zero, $f_{\text{corr}}(t)$ has decayed to almost zero. Therefore, in this case, initial correlations play a negligible role. We find that for specific initial state preparations, the effect of the initial correlations can be largely eliminated.

\section{Summary}

In summary, we have formulated a master equation approach to take into account the effect of a selective system state preparation from
an initial system-environment equilibrium state that has system-environment correlations. Our master equation is valid for weak system-environment coupling.  Two different methods are presented in order to derive the master equation that has an extra term not studied previously.
We have applied our master equation to a variant of the usual spin-boson model. We find that for a collection of two-level atoms coupled to a common environment, the reduced system dynamics can evolve at a faster rate depending on the number of two-level atoms. This finding has implications for quantum control. For instance, in order to preserve a quantum state via dynamical decoupling \cite{LloydDD, largespinmitchell}, we would need to apply the pulses at a faster rate due to the effect studied here. We also considered different initial states to show that the effect of state preparations depends on the actual initial state prepared.

\vspace{0.3cm}
\noindent {\bf Acknowledgment}: J.G. dedicates this work to his late beloved wife Huairui Zhang.

\appendix

\section{Derivation using the time convolutionless approach}
\label{TCLappendix}
Here we sketch an alternative derivation of our master equation using the time convolutionless (TCL) approach. We write the system-environment Hamiltonian as
\begin{equation}
H = H_S + H_B + V = H_0 + \alpha V,
\end{equation}
with $H_S$ the system part, $H_B$ the environment part, and $V$ is the interaction with coupling strength $\alpha$. Working in the interaction picture defined with respect to $H_0$, we can write
\begin{equation}
\frac{d}{dt}\rho_{\text{tot}}(t) = -i\alpha [\widetilde{V}(t), \rho_{\text{tot}}(t)] \equiv \alpha \mathcal{L}(t)\rho_{\text{tot}}(t).
\end{equation}
We re-emphasize that $\rho_{\text{tot}}(t)$ here is in the interaction picture. We now define a projection operator $\mathcal{P}$ such that
\begin{equation}
\mathcal{P}\rho_{\text{tot}} = \text{Tr}_B [\rho_{\text{tot}}] \otimes \rho_B,
\end{equation}
where $\rho_B$ is a reference environment state which we take to be the thermal bath state $\rho_B = e^{-\beta H_B}/Z_B$. The orthogonal projection is given by
\begin{equation}
\mathcal{Q}\rho_{\text{tot}} = \rho_{\text{tot}} - \mathcal{P}\rho_{\text{tot}}.
\end{equation}
The TCL master equation can be written as (see Ref.~\cite{breuerbook} for details)
\begin{equation}
\label{TCLequation}
\frac{\partial}{\partial t} \mathcal{P}\rho_{\text{tot}}(t) = \mathcal{K}(t)\rho_{\text{tot}}(t) + \mathcal{I}(t)\mathcal{Q}\rho_{\text{tot}}(0),
\end{equation}
with the so-called TCL generator given by
\begin{equation}
\mathcal{K}(t) = \alpha \mathcal{P}\mathcal{L}(t)[1 - \Sigma(t)]^{-1}\mathcal{P},
\end{equation}
and
\begin{equation}
\mathcal{I}(t) = \alpha \mathcal{P}\mathcal{L}(t)[1 - \Sigma(t)]^{-1}\mathcal{G}(t)\mathcal{Q},
\end{equation}
where $[1 - \Sigma(t)]^{-1}$ and $\mathcal{G}(t)$ can usually be expanded in powers of the coupling strength $\alpha$. Now, the first term on the right hand side of Eq.~\eqref{TCLequation} leads to the usual second order term in the master equation, so here we focus on the last term in Eq.~\eqref{TCLequation}. The initial state is (see the main text for details)
\begin{equation}
\rho_{\text{tot}}(0) = \ket{\psi}\bra{\psi} \otimes \dfrac{\opav{\psi}{e^{-\beta H}}{\psi}}{Z},
\end{equation}
which can be written as
\begin{equation}
\rho_{\text{tot}}(0) = \rho(0) \otimes \left[ \rho_B^{(0)} + \rho_B^{(1)} + \hdots \right],
\end{equation}
with the environment state expanded in terms of the coupling strength. It is easy to see that $\mathcal{Q}[\rho(0) \otimes \rho_B^{(0)}] = 0$ since $\mathcal{Q} = 1 - \mathcal{P}$. This is the same as for the case of the usual factorizing initial condition where the initial environment state is the thermal bath state. Now, it can be shown that [see Eq.~\eqref{envstatefirstordercorrection} in the main text for details]
\begin{equation}
\rho_B^{(1)} = -\frac{1}{Z_BZ'} \alpha e^{-\beta H_B} E(\beta).
\end{equation}
Using the fact that $\langle E(\beta) \rangle_B$ is generally zero, we obtain $\mathcal{Q}[\rho(0) \otimes \rho_B^{(1)}] =  \rho(0) \otimes \rho_B^{(1)}$. Since we are only concerned with terms up to second order in the coupling strength, we replace $\mathcal{G}$ and $[1 - \Sigma(t)]^{-1}$ by identity. Then, using the definitions of $\mathcal{L}(t)$ and $\mathcal{P}$, we obtain
\begin{align}
&\mathcal{I}(t) \mathcal{Q}\rho_{\text{tot}}(0) = \notag \\
&-\frac{i\alpha^2}{Z_B Z'} \text{Tr}_B [\rho(0) \otimes e^{-\beta H_B} E(\beta), U_0^\dagger (t) F \otimes B U_0(t)] \otimes \rho_B.
\end{align}
Ignoring the fixed environment reference state for simplicity and simplifying the trace, we obtain
\begin{align}
\label{TCLalmostthere}
&\mathcal{I}(t) \mathcal{Q}\rho_{\text{tot}}(0) = \notag \\
&-\frac{i\alpha^2}{Z'} \langle E(\beta)\widetilde{B}(t)\rangle_B [U_S^\dagger (t) [\rho(t), F] U_S(t)],
\end{align}
where $U_S(t) \rho(0) U_S^\dagger (t)$ has been replaced by $\rho(t)$ since any corrections lead to terms of higher order in the coupling strength. Now, note that we are using the interaction picture density matrix in Eq.~\eqref{TCLequation}. Transforming back to the Schrodinger picture, the unitary operators $U_S(t)$ and $U_S^\dagger (t)$ in Eq.~\eqref{TCLalmostthere} are removed. We then obtain the same term due to the initial correlations as Eq.~\eqref{masterequationwithcorr} in the main text.

\section{Derivation of the standard second order master equation}
\label{nocorrmasterequationappendix}

As explained in the main text of the paper, in order to derive the master equation, our task is to simplify
\begin{equation}
\frac{d\rho_{mn}(t)}{dt} = \text{Tr}_{\text{S,B}} \left[ \rho_{\text{tot}}(0) \frac{dX_{nm}(t)}{dt} \right],
\end{equation}
with
\begin{align}
\label{alltermsappendix}
\frac{dX_{nm}(t)}{dt} &= i[H_0^H(t),X_{nm}^H(t)] + i\alpha[\widetilde{V}(t),\widetilde{X}_{nm}(t)] \notag \\
&+ \alpha^2 \int_0^t ds  [[\widetilde{V}(t),\widetilde{X}_{nm}(t)],\widetilde{V}(s)],
\end{align}
for $\rho_{\text{tot}}(0) = \rho(0) \otimes \rho_B$. As before, we find that
$$ \text{Tr}_{\text{S,B}} \left[ \rho_{\text{tot}}(0)i[H_0^H(t), X_{nm}^H(t)]\right] = i\opav{m}{[\rho(t), H_S]}{n}.$$
Also, for this choice of initial state, by the same reasoning as presented in the main text, the contribution of the second term in Eq.~\eqref{alltermsappendix} is zero. So we then need to evaluate
\begin{equation}
\label{tracesystembath}
 \mbox{Tr}_{\text{S,B}} \left[[\rho(t_0) \otimes \rho_B] \int_{0}^t ds [[\widetilde{V}(t),\widetilde{X}_{nm}(t)],\widetilde{V}(s)] \right].
\end{equation}
For simplicity, here we only consider
$$  \mbox{Tr}_{\text{S,B}} \left[[\rho(t_0) \otimes \rho_B] \int_{0}^t ds \widetilde{V}(t)\widetilde{X}_{nm}(t)\widetilde{V}(s) \right].$$
The rest of the terms can be calculated in a similar way. The trace over the bath gives
\begin{align*}
\mbox{Tr}_B \left[\rho_B U_B^\dagger (t) B U_B(t) U_B^\dagger(s) B U_B(s)\right] &=  \\
\langle \widetilde{B}(t) \widetilde{B}(s) \rangle_B.
\end{align*}
The trace over the system is
$$ \mbox{Tr}_S \left[\rho(0) U_S^\dagger (t) F Y_{nm} U_S (t) U_S^\dagger (s) F U_S(s)\right],$$
which simplifies to
\begin{align*}
\mbox{Tr}_S \left[\widetilde{\rho} (t) F Y_{nm} U_S (t,s) F U_S^\dagger (t,s) \right] &= \\
\opav{m}{\bar{F}(t,s)\widetilde{\rho}(t)F}{n},
\end{align*}
with $\bar{F}(t,s) = U_S(t,s)FU^\dagger(t,s)$. We can then make the substitution $\widetilde{\rho}(t) = \rho(t)$. This is justified because the correction gives us terms of higher order in the master equation. By simplifying the other terms [see Eq.~\eqref{tracesystembath}], and putting them all together, we obtain the desired master equation.

\section{Proof that $f_{\text{corr}}(t)$ is real}
\label{proofoffcorrreal}

To actually show that our master equation [see Eq.~\eqref{masterequationwithcorr}] preserves hermiticity, we need to show that $f_{\text{corr}}(t)$ is real. To do this, first recall that
$$ f_{\text{corr}}(t) = \frac{\langle E(\beta) \widetilde{B}(t) \rangle}{Z'}. $$
We can show that $Z'$ is real. We know that
$$Z' = \opav{\psi}{e^{-\beta H_S}}{\psi} - \langle E(\beta) \rangle_B, $$
where
$$ E(\beta) = \int_0^\beta d\lambda \, e^{\lambda H_B} B e^{-\lambda H_B} \opav{\psi}{e^{-\beta H_S} e^{\lambda H_S} F e^{-\lambda H_S}}{\psi}. $$
It follows that
$$E^\dagger(\beta) = \int_0^\beta d\lambda \, e^{-\lambda H_B} B e^{\lambda H_B} \opav{\psi}{ e^{-\lambda H_S} F e^{\lambda H_S} e^{-\beta H_S}}{\psi}. $$
Now, to show that $\opav{\psi}{e^{-\beta H_S}}{\psi}$ is real, we observe that $e^{-\beta H_S} = \sum_i e^{-\beta E_i} \ket{n_i} \bra{n_i} $, where $\ket{n_i}$ are the eigenstates of $H_S$ and $E_i$ the eigenvalues. Then,
$$ \opav{\psi}{e^{-\beta H_S}}{\psi} = \sum_i e^{-\beta E_i} |\langle n_i | \psi \rangle|^2, $$
which is obviously real.

Let us now look at $\langle E(\beta) \rangle_B$. To show this is real, we note that $\langle E(\beta) \rangle_B^* = \langle E^\dagger (\beta) \rangle_B$ and
\begin{align*}
\langle E^\dagger (\beta) \rangle_B = &\int_0^\beta d\lambda \, \text{Tr}_B [\rho_B e^{-\lambda H_B} B e^{\lambda H_B}] \, \times \\
&\opav{\psi}{e^{-\lambda H_S} F e^{\lambda H_S} e^{-\beta H_S}}{\psi}.
\end{align*}
Now perform the variable substitution $\gamma = \beta - \lambda$. It is then straightforward to show that
\begin{align*}
\langle E^\dagger (\beta) \rangle_B = &\int_0^\beta d\gamma \, \text{Tr}_B [\rho_B e^{\gamma H_B} B e^{-\gamma H_B}] \, \times \\
&\opav{\psi}{e^{-\beta H_S} e^{\gamma H_S} F e^{-\gamma H_S}}{\psi},
\end{align*}
which is equal to $\langle E(\beta) \rangle_B$. Therefore, we have shown that $Z'$ is real.

In a similar fashion, we now show that $\langle E(\beta) \widetilde{B}(t) \rangle$ is real. We first note that $\langle E(\beta) \widetilde{B}(t) \rangle_B^* = \langle \widetilde{B}(t) E^\dagger(\beta) \rangle_B$, and it then follows that
\begin{align*}
\langle \widetilde{B}(t) E^\dagger (\beta) \rangle_B = &\int_0^\beta d\lambda \, \text{Tr}_B [\widetilde{B}(t)  e^{-\lambda H_B} B e^{\lambda H_B} \rho_B] \, \times \\
&\opav{\psi}{e^{-\lambda H_S} F e^{\lambda H_S} e^{-\beta H_S}}{\psi}.
\end{align*}
Again using the substitution $\gamma = \beta - \lambda$, we get
\begin{align*}
\langle \widetilde{B}(t) E^\dagger (\beta) \rangle_B = &\int_0^\beta d\gamma \, \text{Tr}_B [\widetilde{B}(t)  \rho_B e^{\gamma H_B} B e^{-\gamma H_B}] \, \times \\
&\opav{\psi}{e^{-\beta H_S} e^{\gamma H_S} F e^{-\gamma H_S}}{\psi},
\end{align*}
which, after using cyclic invariance of the trace operation, can be written as
\begin{align*}
\langle \widetilde{B}(t) E^\dagger (\beta) \rangle_B = &\int_0^\beta d\gamma \, \text{Tr}_B [e^{\gamma H_B} B e^{-\gamma H_B} \widetilde{B}(t) \rho_B] \, \times \\
&\opav{\psi}{e^{-\beta H_S} e^{\gamma H_S} F e^{-\gamma H_S}}{\psi},
\end{align*}
On the other hand,
\begin{align*}
\langle  E(\beta)\widetilde{B}(t) \rangle_B = &\int_0^\beta d\lambda \, \text{Tr}_B [e^{\lambda H_B} B e^{-\lambda H_B} \widetilde{B}(t) \rho_B] \, \times \\
&\opav{\psi}{e^{-\beta H_S} e^{\lambda H_S} F e^{-\lambda H_S}}{\psi}.
\end{align*}
$\langle E(\beta) \widetilde{B}(t) \rangle_B$ is then indeed real. We have therefore shown that $f_{\text{corr}}(t)$ is real.

\section{The exactly solvable large spin pure dephasing model}
\label{puredephasingappendix}

For completeness, we briefly recap the solution of the large spin pure dephasing model with and without initial correlations. We consider the system-environment Hamiltonian to be
\begin{equation}
\label{modelhamiltoniandephasing}
H = H_S + H_B + V,
\end{equation}
with
\begin{align}
H_S = \varepsilon J_z , \, H_B = \sum_k \omega_k b_k^\dagger b_k, \\
V = J_z \sum_k (g_k^*b_k + g_k b_k^\dagger),
\end{align}
which is unitarily equivalent to the system-environment Hamiltonian [see Eq.~\eqref{modelhamiltonian}] considered in the main text with $\Delta = 0$. In order to solve for the dynamics, we first transform to the interaction picture, obtaining
\begin{align}
H_I(t) &= e^{i(H_S + H_B)t} V e^{-i(H_S + H_B)t}, \notag \\
&= J_z \sum_k (g_k^* b_k e^{-i\omega_k t} + g_k b_k^\dagger e^{i\omega_k t} ).
\end{align}
The unitary time evolution operator can then be found using the Magnus expansion to be
\begin{equation}
U_I(t) = \exp \left[ \sum_{i = 1}^{2} A_i(t) \right],
\end{equation}
where
\begin{align}
A_1 &= J_z \sum_k [b_k^\dagger \alpha_k(t) - b_k \alpha_k^*(t)], \\
A_2 &= -iJ_z^2t\Delta(t),
\end{align}
with
\begin{equation}
\alpha_k(t) = \frac{g_k (1 - e^{i\omega_k t})}{\omega_k},
\end{equation}
and
\begin{equation}
\Delta(t) \equiv \frac{1}{t} \sum_k |g_k|^2 \frac{[\sin(\omega_k t) - \omega_k t]}{\omega_k^2}.
\end{equation}
The exact unitary time evolution operator is therefore
\begin{equation}
U(t) = e^{-i\omega_0 J_z t} e^{-iH_B t} U_I(t),
\end{equation}
where
\begin{equation}
U_I(t) = \exp \lbrace J_z \sum_k [b_k^\dagger \alpha_k(t) - b_k \alpha_k^*(t)] - iJ_z^2 t \Delta(t) \rbrace.
\end{equation}

With the time evolution operator available, we can then calculate the reduced density matrix of the system. Expressing the system density in the eigenbasis of $J_z$ (that is, $J_z \ket{n} = n\ket{n}$), we find that
\begin{align}
\label{putininitialstatehere}
[\rho_S(t)]_{mn} = &e^{-i\omega_0 t (m - n)} e^{-i \Delta(t)t(m^2 - n^2)} \times \notag \\
&\text{Tr}_{S,B} [e^{-R_{nm}(t)} P_{nm} \rho(0)],
\end{align}
where $P_{nm} = \ket{n}\bra{m}$, and
\begin{equation}
R_{nm}(t) = (n - m) \sum_k [b_k^\dagger \alpha_k(t) - b_k \alpha_k^*(t)].
\end{equation}
This result is true regardless of the form of the initial state.

Considering first unphysical decorrelated initial states, i.e.,
\begin{equation}
\rho^\text{dir}(0) = \rho_S(0) \otimes \rho_B,
\end{equation}
where $\rho_B = \frac{e^{-\beta H_B}}{Z_B}$ with $Z_B = \text{Tr}_B [e^{-\beta H_B}]$, we find that
\begin{align}
[\rho_S(t)]_{mn} = &[\rho_S(0)]_{mn} e^{-i \omega_0 (m - n) t} e^{-i \Delta(t)(m^2 - n^2)t} \times \notag \\
&e^{-\gamma(t) (m - n)^2 t},
\end{align}
with
\begin{equation}
\gamma(t) = \frac{1}{t} \sum_k |g_k|^2 \frac{[1 - \cos (\omega_k t)]}{\omega_k^2} \coth \left( \frac{\beta \omega_k}{2} \right).
\end{equation}
The factor $e^{-\gamma(t) (m - n)^2 t}$ describes decoherence and the factor $e^{-i \Delta(t)(m^2 - n^2)t}$ describes
the indirect atom-atom interaction induced by the common bath.

Now we consider (physical) correlated initial states of the form
\begin{equation}
\label{preparationviameasurement}
\rho(0) = \frac{1}{Z} P_{\psi} e^{-\beta H_{\text{total}}} P_{\psi}
\end{equation}
with $P_{\psi} = \ket{\psi}\bra{\psi}$ and $Z$ the normalization factor. By using the displaced harmonic oscillator modes
\begin{align}
B_{k,l} = b_k + \frac{lg_k}{\omega_k}, \\
B_{k,l}^\dagger = b_k^\dagger + \frac{lg_k^*}{\omega_k}.
\end{align}
or by using a polaron transformation, it can be shown that \cite{ChaudhryPRA2013}
\begin{eqnarray}
\label{dmcorrelated}
[\rho_S(t)]_{mn}& = &[\rho_S(0)]_{mn} e^{-i\omega_0 (m - n)t} e^{-i \Delta(t)(m^2 - n^2)t}  \nonumber \\
&& \times e^{-\gamma(t)(m - n)^2 t} F_c^{\text{mn}}(t),
\end{eqnarray}
with
\begin{align}
F_c^{\text{mn}}(t) &= \dfrac{\sum_l \left( |\langle l | \psi\rangle |^2 e^{-i\Phi_{nm}^{(l)}(t)} e^{-\beta \omega_0 l}  e^{\beta l^2 \mathcal{C}}\right)}{\sum_l \left( |\langle l | \psi\rangle |^2 e^{-\beta \omega_0 l}  e^{\beta l^2 \mathcal{C}}\right)}, \\
\mathcal{C} &= \sum_k  |g_k|^2/\omega_k, \\
\Phi_{nm}^{(l)} &= (n - m)l \Phi(t), \\
\Phi(t) &= \sum_k \frac{|g_k|^2}{\omega_k^2} \sin (\omega_k t).
\end{align}

\section{Calculations for the effect of initial correlations for different state preparations}
\label{appendixcalculations}

\subsection{$\ket{\psi} = \ket{N/2}$}

The calculation proceeds in a very similar way as before, except that we now have to use
\begin{equation}
e^{-\beta H_S} = e^{\phi J_-} e^{-\phi_z J_z} e^{\phi J_+},
\end{equation}
where
\begin{align}
\phi &= - \frac{\Delta}{\widetilde{\Delta}} \frac{\sinh\left(\frac{\beta \widetilde{\Delta}}{2}\right)}{\mu},  \notag \\
\phi_z &= - 2 \ln \mu, \notag \\
\mu &= \cosh \left( \frac{\beta \widetilde{\Delta}}{2}\right) - \frac{\varepsilon}{\widetilde{\Delta}} \sinh \left( \frac{\beta \widetilde{\Delta}}{2} \right).
\end{align}
We then find that
\begin{align}
&\opav{\psi}{e^{-\beta H_S} e^{\lambda H_S} J_x e^{-\lambda H_S}}{\psi} = \notag \\
&\mu^{N - 1} \frac{N}{2} \left[ \kappa - \frac{\varepsilon \Delta}{\widetilde{\Delta}^2} \cosh\left( \lambda \widetilde{\Delta} - \frac{\beta \widetilde{\Delta}}{2}\right)\right],
\end{align}
with
\begin{equation}
\kappa = -\frac{\Delta}{\widetilde{\Delta}} \sinh\left(\frac{\beta \widetilde{\Delta}}{2}\right) + \frac{\varepsilon \Delta}{\widetilde{\Delta}^2} \cosh\left( \frac{\beta \widetilde{\Delta}}{2}\right).
\end{equation}
Consequently,
\begin{align}
&\frac{\opav{\psi}{e^{-\beta H_S} e^{\lambda H_S} J_x e^{-\lambda H_S}}{\psi}}{Z'} = \notag \\
& \frac{N}{2} \left[ A - \mathcal{B} \cosh \left(\lambda \widetilde{\Delta} - \mathcal{C}\right)\right],
\end{align}
with
\begin{equation}
A = \frac{\kappa}{\mu}, \, \mathcal{B} = \frac{\varepsilon \Delta}{\mu \widetilde{\Delta}^2}, \, \mathcal{C} = \frac{\beta \widetilde{\Delta}}{2}.
\end{equation}
After doing the integration over $\lambda$, we find that
\begin{align}
&\frac{1}{Z'} \langle E(\beta) \widetilde{B}(t) \rangle_B = N \sum_k |g_k|^2 \cos(\omega_k t) \times \notag \\
& \left\lbrace \frac{A}{\omega_k} + \frac{D}{\widetilde{\Delta}^2 - \omega_k^2}\left[\widetilde{\Delta} \coth\left(\frac{\beta \omega_k}{2}\right) - \omega_k \coth\left(\frac{\beta\widetilde{\Delta}}{2}\right) \right]\right\rbrace,
\end{align}
with
\begin{align}
A = -\frac{\Delta}{\widetilde{\Delta}} \frac{\widetilde{\Delta} - \varepsilon \coth(\beta \widetilde{\Delta}/2)}{\widetilde{\Delta} \coth(\beta\widetilde{\Delta}/2) - \varepsilon}, \\
D = - \frac{\varepsilon\Delta/\widetilde{\Delta}}{\widetilde{\Delta} \coth(\beta\widetilde{\Delta}/2) - \varepsilon}.
\end{align}

\subsection{Each spin prepared in a coherent superposition}
We perform our calculations in the rotated frame, where we have,
\begin{equation}
H^R = H^R_S + H_B + V^R,
\end{equation}
with
\begin{align}
H^R_S = \varepsilon_r J_z + \Delta_r J_x, \, H_B = \sum_k \omega_k b_k^\dagger b_k, \\
V^R = J_z \sum_k (g_k^*b_k + g_k b_k^\dagger).
\end{align}
Our initial state is now $\ket{\psi^R}$, which is an eigenstate of $J_z$ with eigenvalue $\frac{N}{2}$.
We first note that
\begin{equation}
e^{\lambda H_S^R} J_z e^{-\lambda H_S^R} = a_x J_x + a_y J_y + a_z J_z,
\end{equation}
where
\begin{align}
a_x &= \frac{\varepsilon_r \Delta_r}{\widetilde{\Delta}_r^2} \left[ 1 - \cosh (\lambda \widetilde{\Delta}_r)\right], \notag \\
a_y &= -\frac{i\Delta_r}{\widetilde{\Delta}_r} \sinh(\lambda \widetilde{\Delta}_r), \notag \\
a_z &= \frac{\varepsilon_r^2 + \Delta_r^2 \cosh(\lambda \widetilde{\Delta}_r)}{\widetilde{\Delta}_r^2}.
\end{align}
Our initial state is now $\ket{\psi^R} = \ket{N/2}$. To proceed further, we use the identity
\begin{equation}
e^{-\beta H_S} = e^{\phi J_-} e^{-\phi_z J_z} e^{\phi J_+},
\end{equation}
where
\begin{align}
\phi &= - \frac{\Delta_r}{\widetilde{\Delta}_r} \frac{\sinh\left(\frac{\beta \widetilde{\Delta}_r}{2}\right)}{\mu},  \notag \\
\phi_z &= - 2 \ln \mu, \notag \\
\mu &= \cosh \left( \frac{\beta \widetilde{\Delta}_r}{2}\right) - \frac{\varepsilon_r}{\widetilde{\Delta}_r} \sinh \left( \frac{\beta \widetilde{\Delta}_r}{2} \right).
\end{align}
Proceeding the same way as before, we can find that
\begin{align}
&\langle\psi^R | e^{-\beta H_S^R} e^{\lambda H_S^R} J_z e^{-\lambda H_S^R} | \psi^R \rangle = \notag \\
&= \mu^{N - 1} \frac{N}{2} \left[ \kappa + \frac{\Delta_r^2}{\widetilde{\Delta}_r^2} \cosh \left(\lambda \widetilde{\Delta}_r - \frac{\beta \widetilde{\Delta}_r}{2} \right)\right],
\end{align}
where
\begin{equation}
\kappa = -\frac{\varepsilon_r}{\widetilde{\Delta}_r}\sinh\left(\frac{\beta\widetilde{\Delta}_r}{2}\right) + \frac{\varepsilon_r^2}{\widetilde{\Delta}_r^2} \cosh\left(\frac{\beta \widetilde{\Delta}_r}{2}\right).
\end{equation}
It can then be shown that
\begin{align}
&\frac{1}{Z'} \langle E(\beta) \widetilde{B}(t) \rangle_B = N \sum_k |g_k|^2 \cos(\omega_k t) \times \notag \\
& \left\lbrace \frac{A}{\omega_k} + \frac{D}{\widetilde{\Delta}_r^2 - \omega_k^2}\left[\widetilde{\Delta}_r \coth\left(\frac{\beta \omega_k}{2}\right) - \omega_k \coth\left(\frac{\beta\widetilde{\Delta}_r}{2}\right) \right]\right\rbrace,
\end{align}
with
\begin{align}
A = -\frac{\varepsilon_r}{\widetilde{\Delta}_r} \frac{\widetilde{\Delta}_r - \varepsilon_r \coth(\beta \widetilde{\Delta}_r/2)}{\widetilde{\Delta}_r \coth(\beta\widetilde{\Delta}_r/2) - \varepsilon_r}, \\
D = \frac{\Delta_r^2}{\widetilde{\Delta}_r^2} \frac{1}{\coth(\beta\widetilde{\Delta}_r/2) - \varepsilon_r/\widetilde{\Delta}_r}.
\end{align}


\begin{thebibliography}{99}
\bibitem{footnoteonV} It is simple to extend to treatment to the case of $V = \sum_j F_j \otimes B_j$.
\end{thebibliography}
\end{document}